\documentclass[twocolumn,tighten,iop,times,twocolappendix]{aastex631}
\usepackage{graphicx}
\graphicspath{{Plots/}}
\usepackage{amsmath}
\usepackage{farhan-defs}
\usepackage[super]{nth}

\usepackage{bigints}
\usepackage[fulladjust]{marginnote}

\usepackage{appendix}

\hypersetup{colorlinks=true, citecolor=blue, filecolor=magenta,urlcolor=blue, linkcolor=blue}

\begin{document}


\title{\sc The Evolving Effect Of Cosmic Web Environment On Galaxy Quenching}


\author[0000-0002-0072-0281]{Farhanul Hasan}
\affiliation{Department of Astronomy, New Mexico State University, Las Cruces, NM 88003, USA}

\author[0000-0002-1979-2197]{Joseph N. Burchett}
\affiliation{Department of Astronomy, New Mexico State University, Las Cruces, NM 88003, USA}

\author{Alyssa Abeyta}
\affiliation{Department of Astronomy, New Mexico State University, Las Cruces, NM 88003, USA}

\author{Douglas Hellinger}
\affiliation{Department of Physics, University of California, Santa Cruz, CA 95064, USA}

\author[0000-0001-8057-5880]{Nir Mandelker}
\affiliation{Centre for Astrophysics and Planetary Science, Racah Institute of Physics, The Hebrew University, Jerusalem 91904, Israel}

\author[0000-0001-5091-5098]{Joel R. Primack}
\affiliation{Department of Physics, University of California, Santa Cruz, CA 95064, USA}

\author[0000-0003-4996-214X]{S. M. Faber}
\affiliation{University of California Observatories and Department of Astronomy and Astrophysics, \\
University of California, Santa Cruz, 1156 High Street, Santa Cruz, CA 95064, USA}

\author[0000-0003-3385-6799]{David C. Koo}
\affiliation{University of California Observatories and Department of Astronomy and Astrophysics, \\
University of California, Santa Cruz, 1156 High Street, Santa Cruz, CA 95064, USA}

\author[0000-0003-0549-3302]{Oskar Elek}
\affiliation{Department of Computational Media, University of California, 1156 High Street, Santa Cruz, CA 95064, USA}

\author[0000-0002-6766-5942]{Daisuke Nagai}
\affiliation{Department of Physics, Yale University, New Haven, CT 06520, USA}


 \correspondingauthor{Farhanul Hasan} 
 \email{farhasan@nmsu.edu}

 \shorttitle{Cosmic Web Environment And Galaxy Quenching} \shortauthors{Hasan et al.}


\begin{abstract}

We investigate how cosmic web structures affect galaxy quenching in the IllustrisTNG (TNG100) cosmological simulations by reconstructing the cosmic web within each snapshot using the {\disperse} framework.
We measure the comoving distance from each galaxy with stellar mass {\minms} to the nearest node ({\dnode}) and the nearest filament spine ({\dfil}) to study the dependence of both median specific star formation rate ({\medssfr}) and median gas fraction ({\medfgas}) on these distances. 
{\it We find that the {\medssfr} of galaxies is only dependent on cosmic web environment at $z<2$}, with the dependence increasing with time. 
At $z\leq0.5$, {\lowmsrange} galaxies are quenched at ${\dnode}\lesssim1$~Mpc, and have significantly-suppressed star formation at ${\dfil}\lesssim1$~Mpc, trends driven mostly by satellite galaxies.
At $z\leq1$, in contrast to the monotonic drop in {\medssfr} of {\mslow} galaxies with decreasing {\dnode} and {\dfil}, {\highmsrange} galaxies -- both centrals and satellites -- experience an upturn in {\medssfr} at ${\dnode}\lesssim0.2$~Mpc.
Much of this cosmic web dependence of star formation activity can be explained by an evolution in {\medfgas}. 
Our results suggest that in the past $\sim$10~Gyr, low-mass satellites are quenched by rapid gas stripping in dense environments near nodes and gradual gas starvation in intermediate-density environments near filaments. At earlier times, cosmic web structures efficiently channeled cold gas into most galaxies.
State-of-the-art ongoing spectroscopic surveys such as SDSS and DESI, as well as those planned with the Subaru Prime Focus Spectrograph, {\it JWST}, and {\it Roman}, are required to test our predictions against observations.

\end{abstract}

\received{2023 Mar 13}
\revised{2023 Apr 21}
\accepted{2023 Apr 23}

 \keywords{Cosmic web (330), Large-scale structure of the universe (902), Galaxy quenching (2040), Galaxy evolution (594), Intergalactic filaments (811), Magnetohydrodynamical simulations (1966)}

\section{Introduction}

In the standard cosmological model, structure formation in the universe occurs at vastly different scales. Galaxies form stars within tens of kpc and grow inside dark matter (DM) halos that can be two orders of magnitude larger. At even larger scales, galaxies and their DM halos are embedded within an intricate network of strand-like filaments, diffuse sheets, dense nodes, and underdense voids, which is termed the ``cosmic web'' \citep[e.g.,][]{Bond96, Springel05}. 
While the cosmic web has been studied on both theoretical and observational grounds for decades, it remains one of the major outstanding questions in astrophysics whether and how the large-scale cosmic web environment influences the formation and evolution of galaxies.

A chief question in galaxy evolution is how star formation activity proceeds in galaxies and how it ceases, i.e., how quenching occurs. It has long been known that quenching depends on internal mechanisms characterized by the stellar mass {\ms} or halo mass $M_{\mathrm{vir}}$ \citep[e.g.,][]{Brinchmann04,Cattaneo06,Williams09,Peng10,Darvish16}, such that galaxy-scale processes, including supernovae (SNe) and Active Galactic Nuclei (AGN) feedback, can regulate and curtail star formation activity. 
A widely adopted theoretical viewpoint posits that galaxies in halos with mass $\log(M_{\mathrm{vir}}/{\Msun})\gtrsim11.5-12$ can form stable virial accretion shocks and, therefore, a hot, hydrodynamically stable circumgalactic medium (CGM) that suppresses accretion of cold gas to the interstellar medium (ISM) necessary for star formation \citep[e.g.,][]{DB06,Keres05,Keres09,Dekel09,Stern20,Stern21}. Lower-mass galaxies (typically with $\log({\ms}/{\Msun})\!<\!10$), 
especially at high redshifts ($z\!\gtrsim\!2$), lack this ability to form a stable hot CGM and ``self-quench'' \citep[e.g.,][]{Croton06,GD12}.

External processes as characterized by their environment have also emerged as crucial factors in determining how galaxies quench 
\citep[e.g.,][]{Elbaz07,Peng12,Eardley15,Moutard18,Bluck20}. However, the exact nature of the relationship between quenching and environment, and what physical mechanisms manifest this relationship, is a topic of widespread debate. 
In the hot, dense halos of galaxy groups and clusters, hydrodynamical interactions between the halo medium and satellite galaxies -- most notably ram pressure stripping \citep[e.g.,][]{BM15,Boselli22} -- or tidal interactions between separate galaxies or between galaxies and the halo \citep[e.g.,][]{BG06,Marasco16} can remove the star-forming ISM of a galaxy. Over longer timescales, gas accretion onto the ISM can be halted, either due to lack of accretion from the intergalactic medium (IGM) to the CGM or from the CGM to the ISM via strangulation or starvation \citep[e.g.,][]{Larson80,BM00,Peng15}.

The cosmic web itself has also been invoked in models of galaxy quenching. \citet{AC19} proposed that ``cosmic web detachment,'' wherein galaxies are detached from cold gas-supplying primordial filaments, can explain much of the observed quenching phenomena across time. 
\citet{Song21} suggested that close to the edges of filaments there is coherent, high angular momentum supply of gas to the outer parts of halos, which prevents an efficient transfer of gas from the outer halo to galactic centers -- ultimately quenching these galaxies \citep[see also][]{PR20,Renzini20}.
\citet{Pasha22} found that cosmological accretion shocks at $z\!\sim\!2-5$ can produce a hot ($T\!>\!10^{6}$~K) IGM at the edge of sheets, which can quench low-mass centrals at these epochs, as shocks around filaments, groups, and clusters  can at lower redshifts \citep[e.g.,][]{Birnboim16,Zinger18,Li23}.

Studies of the connection between galaxy quenching and the cosmic web in the past decade have yielded mixed results. While many observational studies have found that passive or quenched galaxies are typically located near nodes and filaments \citep[e.g.,][]{Kuutma:2017aa,Kraljic18, Laigle18, Winkel21}, some have shown that proximity to cosmic web filaments can also enhance star formation in galaxies \citep[e.g.,][]{Darvish14,Vulcani19}. Cosmological hydrodynamical simulations have also provided an inconclusive picture. In the IllustrisTNG simulations \citep[][]{TNGDR19}, \citet{Malavasi22} found that the specific star formation rate ($\mathrm{sSFR} \!=\! \mathrm{SFR}/{\ms}$) of galaxies is generally reduced with proximity to nodes and filaments at $z\!=\!0$.
\citet{Xu20} found in the EAGLE simulations \citep[][]{Schaye15} 
a characteristic stellar mass ($\log({\ms}/{\Msun})\!\sim\!10.5$) below which galaxies have lower sSFR in nodes than in filaments and above which this dependence vanishes. Both \citet{Kotecha22} and \citet{Zheng22} 
reported evidence, instead, of filaments increasing star formation activity or at least delaying quenching. 
Therefore, consensus is yet to be reached on the impact of cosmic web environment on galaxy quenching and how this varies with stellar mass and redshift.

In this paper, we employ the IllustrisTNG cosmological simulations to study the impacts of cosmic web environment, particularly the proximity to filaments and nodes, on star formation and gas content in galaxies across cosmic time. We reconstruct the cosmic web in IllustrisTNG using the topologically-motivated {\disperse} framework \citep{disperse1,disperse2}. This is the first study of the dependence of star formation quenching on the cosmic web in the TNG100-1 run across many different redshift snapshots. \citet{Malavasi22} performed a similar analysis of the TNG300-1 run at $z=0$.

This paper is organized as follows. In Section~\ref{sec:data}, we describe the simulation data used in this work and methods of reconstructing the cosmic web. We present our results
in Section~\ref{sec:results}. We discuss the physical interpretations of our results and propose observational tests in Section~\ref{sec:discuss}, and conclude in Section~\ref{sec:conclusion}.
We adopt the {\it Planck 2015} cosmology \citep{Planck15}, with $H_{0}=67.74$ {\kmsmpc}, $\Omega_{\mathrm{M},0} = 0.3089$, and $\Omega_{\Lambda,0} = 0.6911$. All distances are quoted in comoving units, unless stated otherwise.


\section{Data and Methods}
\label{sec:data}

\subsection{TNG Simulations}
\label{sec:tngdata}

We analyzed outputs from the IllustrisTNG magneto-hydrodynamical cosmological simulations, which use the AREPO moving-mesh hydrodynamics code \citep{Springel10} to simulate the evolution of gas, stars, DM, and black holes (BH) from the early universe ($z=127$) to the present day ($z=0$). The public data release of the simulations was presented in \citet{TNGDR19}, while introductory results were presented in \citet{Pillepich18}, \citet{Nelson18}, \citet{Springel18}, \citet{Marinacci18}, and \citet{Naiman18}.
In particular, we make use of TNG100-1, the highest resolution run of the TNG100 simulation, which has a box size of $\sim$110.7 comoving Mpc per side, minimum baryonic and DM particle mass of $\sim\!1.4 \!\times\! 10^{6}~{\Msun}$ and $\sim7.5 \!\times\! 10^{6}~{\Msun}$ respectively, a {\it Planck 2015} cosmology \citep{Planck15}, and $1820^3$ initial DM particles. 
While TNG300-1 provides greater statistics of galaxies and cosmic structures with $\approx\!20$ times the volume of TNG100-1, it has $\approx\!1/8$ the particle mass resolution. On the other hand, TNG50-1 provides $\approx\!16\times$ greater particle mass resolution than TNG100-1, but has $\approx\!1/10$ the volume. 

We obtain galaxy data for all 100 snapshots of the TNG100-1 simulation (hereafter TNG) from the online data repository\footnote{\url{https://www.tng-project.org/data/}} \citep{TNGDR19}. 
In each snapshot, ``Group'' catalogs are constructed using the friends-of-friends (FoF) substructure identification algorithm, while the {\sc Subfind} algorithm and searches for gravitationally bound objects in each FoF group representing either subhalos or the main (host) halo \citep{Springel01,Dolag09}. We make use of both the group and subhalo catalogs to identify halos and galaxies, respectively.

\begin{figure*}[htbp]
\vspace{-10pt}
\gridline{\fig{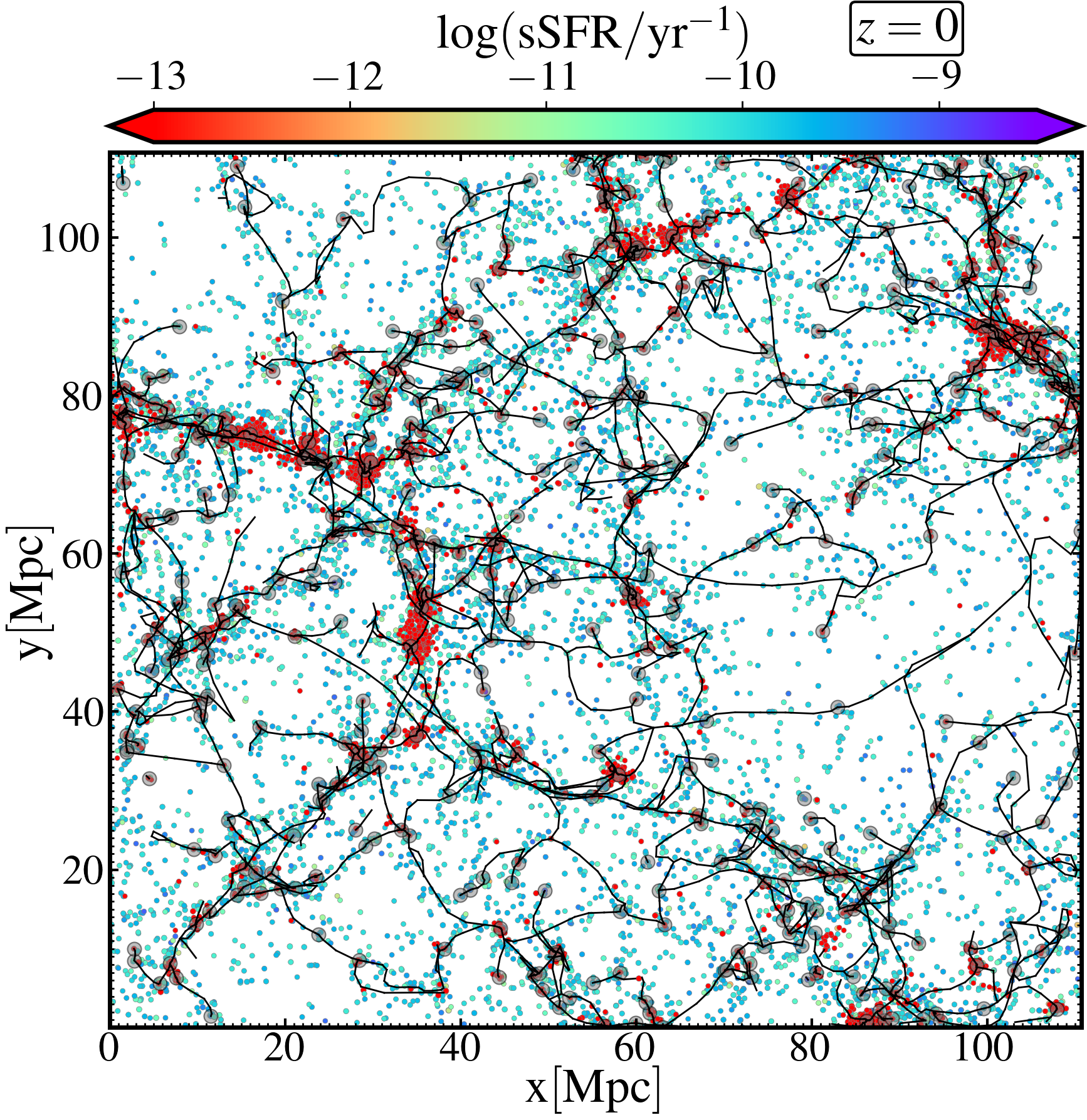}{0.33\textwidth}{}
\hspace{-5pt}
\fig{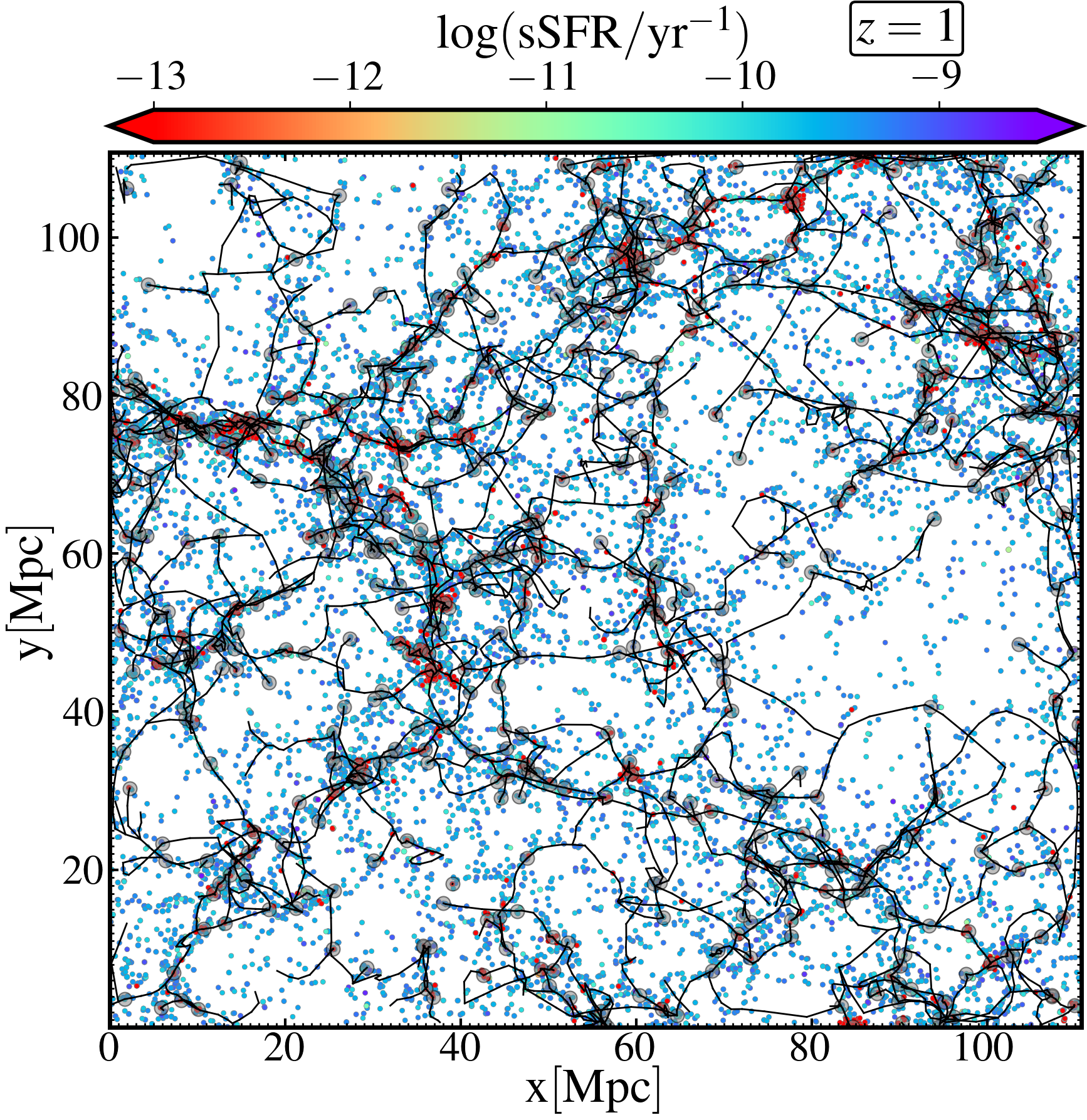}{0.33\textwidth}{}
\hspace{-5pt}
\fig{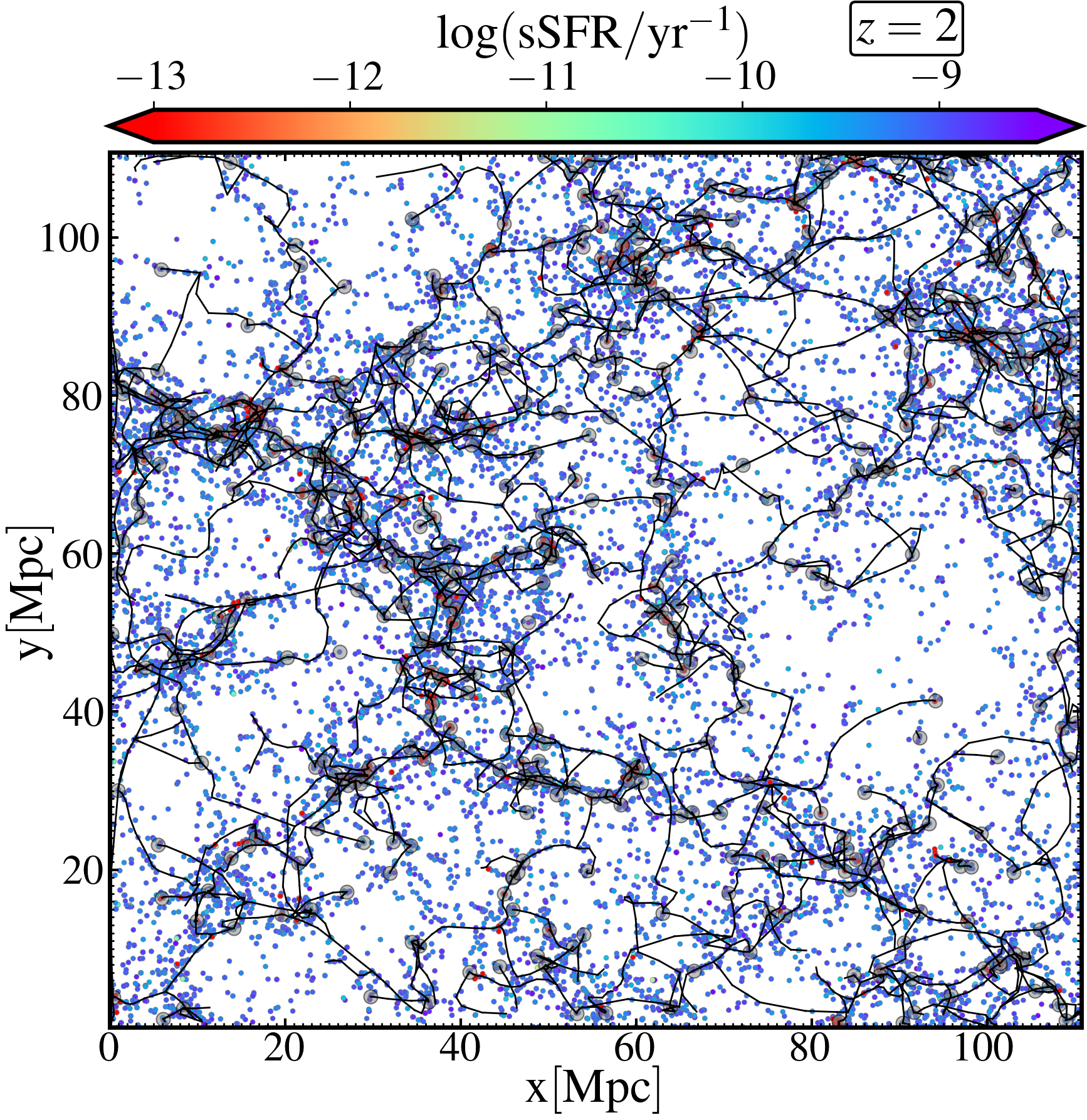}{0.33\textwidth}
{}
}
\vspace{-25pt}
\caption{2D visual representation of galaxies in the TNG100 simulation and cosmic web structures identified by {\disperse} at $z=0$  (left), $z=1$ (middle), and $z=2$ (right). 
In each panel, filament spines are represented by black curves, and nodes are represented by semi-transparent grey circles, while galaxies are represented by scatter points sized by the stellar mass and color-coded by specific star formation rate.
The same x-y projection of a slice 37 Mpc thick ($\sim1/3$ of the total box width) is shown for each redshift. 
The global star formation activity declines considerably from higher to lower redshift.
Quiescent galaxies at lower redshifts appear to be clustered closer to nodes and filaments.
}
\label{fig:vis}
\end{figure*}

For each snapshot, 
we set a minimum stellar mass of $\log({\ms}/{\Msun}) \!=\! 8$, which corresponds to a typical minimum observable stellar mass of galaxies in the (nearby) universe and also ensures that the galaxies are well-resolved with at least about 100 stellar particles in them. Similarly, we set a minimum halo mass -- corresponding to the mass enclosed in a sphere whose mean density is 200 times the critical density of the Universe -- of $\log({\mh}/{\Msun}) \!=\! 9$ to ensure that each of the galaxies in our catalog resides inside halos that are well-resolved with at least 100 DM particles. These criteria yielded $\sim$50,000 galaxies at $z\!=\!0$ and $\sim$11,000 galaxies at $z\!=\!5$. From these catalogs, we obtained the galaxy comoving position, star formation rate (SFR), stellar mass ({\ms}), halo mass ({\mh}), halo virial radius ({\Rh}; comoving radius at which {\mh} is calculated), and mass of all gas gravitationally bound to a subhalo ({\mgas}). 
Hereafter, we refer to subhalos as galaxies and groups as halos.


\subsection{Reconstructing the Cosmic Web with {\disperse}}
\label{sec:disperse}

Next, we apply the Discrete Persistent Structures Extractor ({\disperse}) algorithm \citep{disperse1,disperse2} to find cosmic web filaments and nodes in each TNG snapshot where at least 10,000 galaxies matched our selection criteria above. {\disperse} identifies the topology of space in any given volume based on an input distribution of discrete tracers, which in our case are the spatial locations of all galaxies matching our selection criteria above. 
To do this, it computes the density field from the inputs, using the Delaunay Tessellation Field Estimator \citep[DTFE; ][]{SV00}, wherein the entire volume is divided into tetrahedrons, with the positions of individual galaxies as vertices.
During the tessellation, the density field at the position of each vertex of the tessellation is smoothed by averaging it with its two nearest neighbors. 
This is done in order to minimize contamination by shot noise and the detection of small-scale spurious features \citep[see, e.g.,][]{Malavasi22}. 
{\disperse} calculates the gradient of the density field and identifies critical points where the gradient is zero. These correspond to the voids (minima), saddle points, and nodes (maxima) of the density field. Filaments consist of a series of segments connecting maxima to other critical points.

For each topologically significant pair of critical points, {\disperse} computes the persistence, which is defined as the ratio of the density value at the two critical points.
Persistence is a simple measure of how robust topological structures, i.e., the identified critical points and filament segments, are to local variations, in this case, of the density field measured from input galaxy positions. This sets the effective significance level of the detected filaments and allows us to quantify the effect of shot noise in the input data. 
For our fiducial run, we choose a persistence threshold of $3\sigma$, which has been known to eliminate most spurious filamentary features in the TNG simulations \citep[e.g.,][]{GE20a}. By experimenting with cuts of $4\sigma$ and $5\sigma$, we find that these miss fainter structures, but, nonetheless, do not significantly alter our results.

In addition, we choose to apply a smoothing to the position of the segments of the filamentary skeleton by averaging the initial positions of the extrema of a segment with those of the extrema of contiguous segments. In essence, the skeleton is smoothed by keeping the critical points fixed and averaging the coordinates of each point along a filament with that of its two neighbors. 
This is done to reduce sharp and/or unphysical shapes of filament segments caused by shot noise
We apply one level of smoothing and find that increasing the amount of smoothing by one level or removing this smoothing does not have a significant effect on our statistical results.

Furthermore, we experimented with varying the minimum stellar mass cut of our galaxy catalog and cosmic web reconstruction, using cuts of $\log({\ms}/{\Msun})\!\geq\!9$ and {\highmsrange}. These are more realistic minimum masses to compare to large observational surveys such as the Sloan Digital Sky Survey \citep[SDSS; e.g.,][]{Strauss02} DR17 
(see, e.g., \cite{Wilde23}, and Section~\ref{sec:obs} of this paper).
However,  the sharp drop in the number of galaxies in TNG with these higher masses yielded increasingly fewer input tracers for {\disperse}, 
which resulted in far fewer identified filaments and nodes than in our fiducial {\minms} cut, and particularly very few short filaments (with length $<$1 Mpc).
These higher mass cuts could therefore bias our results towards more prominent cosmic web features and longer filaments.
In practice, varying the minimum {\mh} cut was mostly degenerate with varying the minimum {\ms} cut.

A 2D visual representation of the {\disperse}-identified filaments and nodes superimposed on the distribution of galaxies in TNG is shown in Fig.~\ref{fig:vis}. 
The three panels show $x-y$ projections in a 37 Mpc thick slice (corresponding to $\sim1/3$ of the total thickness) at $z\!=\!0$ (left), $z\!=\!1$ (middle), and $z\!=\!2$ (right).
In each panel, the filament spines and nodes are represented by black curves and grey circles, respectively, while galaxies are represented by scatter points, with sizes proportional to {\ms} and color-coding by sSFR.

From this visualization, we can 
qualitatively assess the spatial distribution of star formation activity in galaxies with respect to cosmic web nodes and filaments. At any redshift, higher sSFR galaxies are located throughout the volume, whereas lower sSFR galaxies are almost always located close to a filament spine or a node. From higher to lower redshift (right to left panel), there is a clear decline in global star formation activity which reflects the decline in cosmic star formation rate density after the so-called ``cosmic noon'' thoroughly chronicled in observations \citep[e.g.,][]{MD14}. The number of massive quiescent galaxies increases considerably from $z\!=\!2$ to $z\!=\!1$ and even more prominently from $z\!=\!1$ to the present day.
A rough qualitative visual check showed that virtually all $\mathrm{SFR}\!=\!0$ galaxies at low redshift, regardless of mass, live near nodes and/or filaments (we quantify a galaxy's proximity to these cosmic web structures below). 


\subsection{Defining Distances}

To quantitatively study the relationship between the physical properties of a galaxy and its cosmic web environment, we measure two different distances for each galaxy at each snapshot: {\dnode} -- the comoving Euclidean distance from the center of a galaxy to the center of the nearest identified node, and {\dfil} -- the comoving transverse distance from the center of a galaxy to the nearest identified filament spine. We chose these cosmic web-centric distance characterizations as they are similar to those of \citet{Welker20} and \citet{Malavasi22}, among others. 
Depending on the physical mechanisms affecting star formation quenching, use of physical distances or other parameters such as local gas density, pressure, angular momentum, etc. are also viable options for the study of the dependence of galaxy properties on the cosmic web environment.
In the following, we investigate how star formation quenching and gas reservoirs of TNG galaxies depend on {\dnode} and {\dfil} at different masses and redshifts.


\section{Results}
\label{sec:results}

\subsection{Star Formation And Cosmic Web Environment}
\label{sec:sf}

We first investigate the relationship between star formation activity in galaxies and their proximity to cosmic web structures. In each snapshot, we divide galaxies into three different stellar mass ranges -- {\lowmsrange}, {\midmsrange}, and {\highmsrange} -- and into seven bins
of the distances {\dnode} and {\dfil}. These bins are chosen such that each has an equal number of galaxies. We measure the median sSFR, {\medssfr}, for each bin of {\dnode} and {\dfil}. Experimenting with different numbers of bins, we find the overall results to be insensitive to the number of bins.

\begin{figure*}
\gridline{\fig{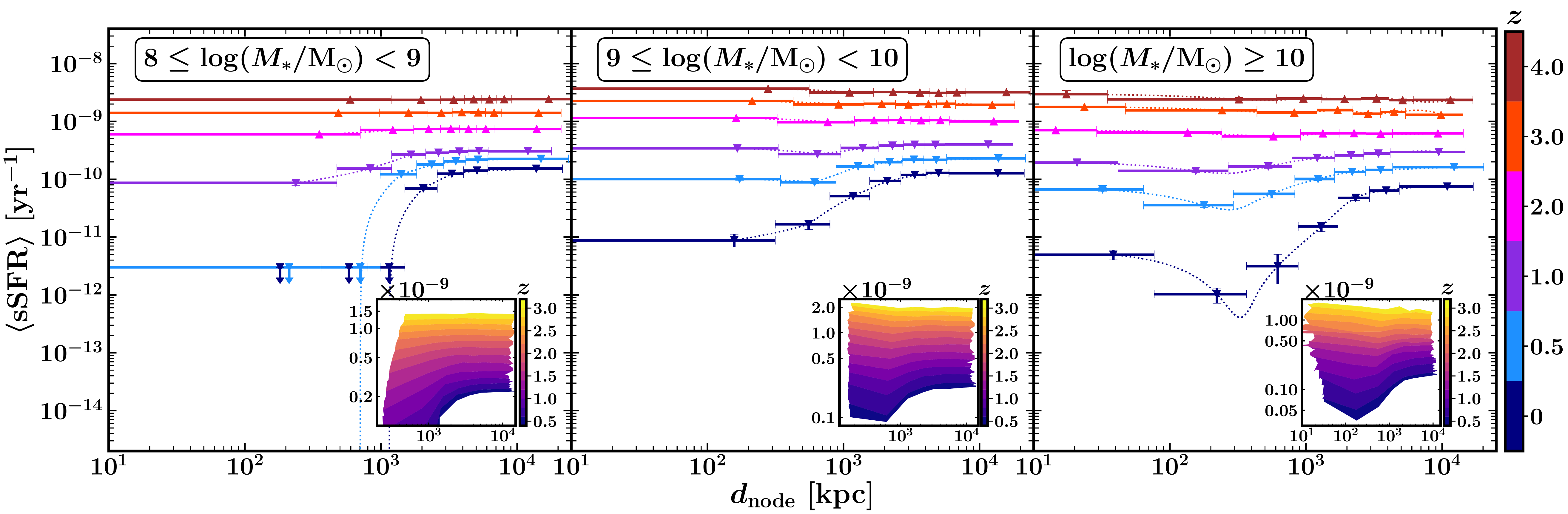}{0.99\textwidth}{}
} \vspace{-28pt}
\gridline{\fig{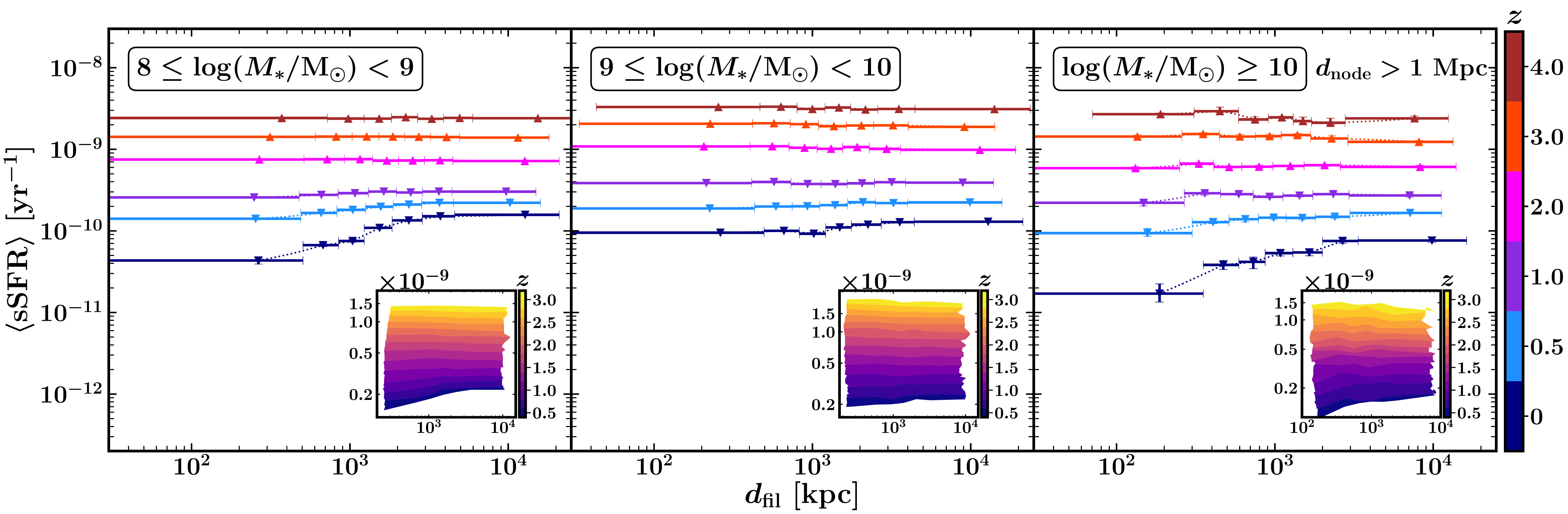}{0.99\textwidth}{}
} \vspace{-25pt}
\caption{
The median sSFR as a function of distance to the nearest node (top row) and filament spine (bottom row) 
for galaxies with {\lowmsrange} (left panels), {\midmsrange} (middle panels), and {\highmsrange} (right panels). The data points and curves are color-coded by redshift as indicated by the discrete color-bars on the right. 
Each panel contains an inset which is a continuous color contour plot showing the {\medssfr}-{\dnode} or {\medssfr}-{\dfil} relationship for a larger number of intermediate redshifts, to help locate the redshift at which a distance-dependence disappears (see text). Only galaxies with ${\dnode}>1$~Mpc are included for the filament-centric relationships to mitigate the potential halo-centric effects of nearby clusters and groups. Note that some points are shown as upper limits on {\medssfr}. Star formation activity is dependent on {\dnode} and (to a lesser extent) on {\dfil} only at lower redshifts, while this dependence disappears at $z\geq 2$. 
}
\label{fig:medssfr}
\end{figure*}

The results of these binned statistics are presented in Fig.~\ref{fig:medssfr}, the top row showing {\medssfr} as a function of {\dnode} and the bottom row {\medssfr} as a function of {\dfil}, and each column represent a different mass range. 
The {\medssfr} are color-coded by redshift;
vertical error bars represent $\pm1\sigma$ bootstrapped errors on {\medssfr} in each {\dnode} or {\dfil} bin, and horizontal error bars represent the width of the bin. Dotted curves represent simple spline interpolations to the {\medssfr}-{\dnode} and {\medssfr}-{\dfil} relations. For a finer look at some intermediate redshifts, each panel also contains an inset showing a color contour plot for each snapshot between $z\!=\!0.5$ and $z\!=\!3$. For certain bins, we show an upper limit on {\medssfr}, corresponding to SFR$=\!10^{-2.5}~{\Msunyr}$, which is the minimum resolvable SFR (averaged over 200 Myr) in TNG due to stochastic star formation with a minimum star particle mass \citep[see][]{Terrazas20}. For example, for {\lowmsrange} galaxies, the maximum upper limit in sSFR is $10^{-11.5}~\mathrm{yr}^{-1}$, considering the most massive galaxies in this mass range.

In order to separate the effect of nearby nodes from that of filaments alone (because many galaxies that are close to filament spines are also close to nodes), we only considered galaxies that are ${\dnode}\!>\!1$~Mpc away from the nearest node for the bottom row of Fig.~\ref{fig:medssfr}. The choice of 1 Mpc is motivated by two considerations. 1) This is slightly higher than the virial radius {\Rh} of the most massive galaxy cluster in TNG100. 
Therefore, this ensures that we remove possible halo-centric effects of nearby clusters and groups (which reside in nodes) and isolates the effect of nearby filaments.
2) This is four times the scale radius of the number density profile of galaxies in filaments derived for TNG300 by \citet{GE20a} and corresponds to the width containing almost all of the matter inside of filaments.
We vary this cut to be ${\dnode}\!>\!0.5$,~$1.5$, and $2$ Mpc as well and find that only galaxies at $z\!\lesssim1\!$ with ${\dfil}\!<\!1$~Mpc show noticeable changes to our quantitative results, while the qualitative results described below remain unchanged.
In essence, this {\dnode} cut ensures that only galaxies at intermediate to high, rather than extremely high local densities, are considered for the {\dfil} analysis (see \citet{Burchett20} for an example of how cosmic matter densities relate to different filamentary environments).

First, examining the node-centric relationships, we find that {\dnode} is strongly correlated with quenching of star formation in galaxies of all masses at low redshifts ($z\!\lesssim\!0.5$). For {\lowmsrange} galaxies at $z\!\leq\!0.5$, {\medssfr} vanishes at ${\dnode}\!\lesssim\!1$~Mpc. The increase from low to high {\dnode} is much more gradual at $\!z=\!1$ ($\approx\!3\times$ from $\langle{\dnode}\rangle\!\sim\!0.2$~Mpc to $\langle{\dnode}\rangle\!\sim\!15$~Mpc). At $z\!\geq\!2$ however, there is virtually no dependence of {\medssfr} on {\dnode}. 
We note that our results at $z\!=\!0$ are broadly in agreement with those of \citet{Geha12}, who found that in SDSS DR8, almost all quenched galaxies with $7\leq\log(\ms/\Msun)\leq9$ are found within $\sim$1.5~Mpc of a massive host.

For intermediate-mass {\midmsrange} galaxies, the trends are similar in that there is a large $\sim$1 dex rise in {\medssfr} from the lowest to the highest {\dnode} bin at $z\!=\!0$, a much smaller rise at $z\!=\!0.5$, and effectively no {\dnode} dependence at $z\!\geq\!1$. The lack of {\dnode}-dependence of {\medssfr} at $z\!>\!1$ is also seen for high-mass {\highmsrange} galaxies, but, interestingly, these galaxies do not show a monotonic increase in {\medssfr} with {\dnode} at lower redshifts. In fact,  following a decline with decreasing {\dnode} at ${\dnode}\!\gtrsim\!0.2$~Mpc, there is an {\it upturn} in {\medssfr} at ${\dnode}\!\lesssim\!0.2$~Mpc.

\begin{figure*}
\vspace{-10pt}
\centering
\gridline{\fig{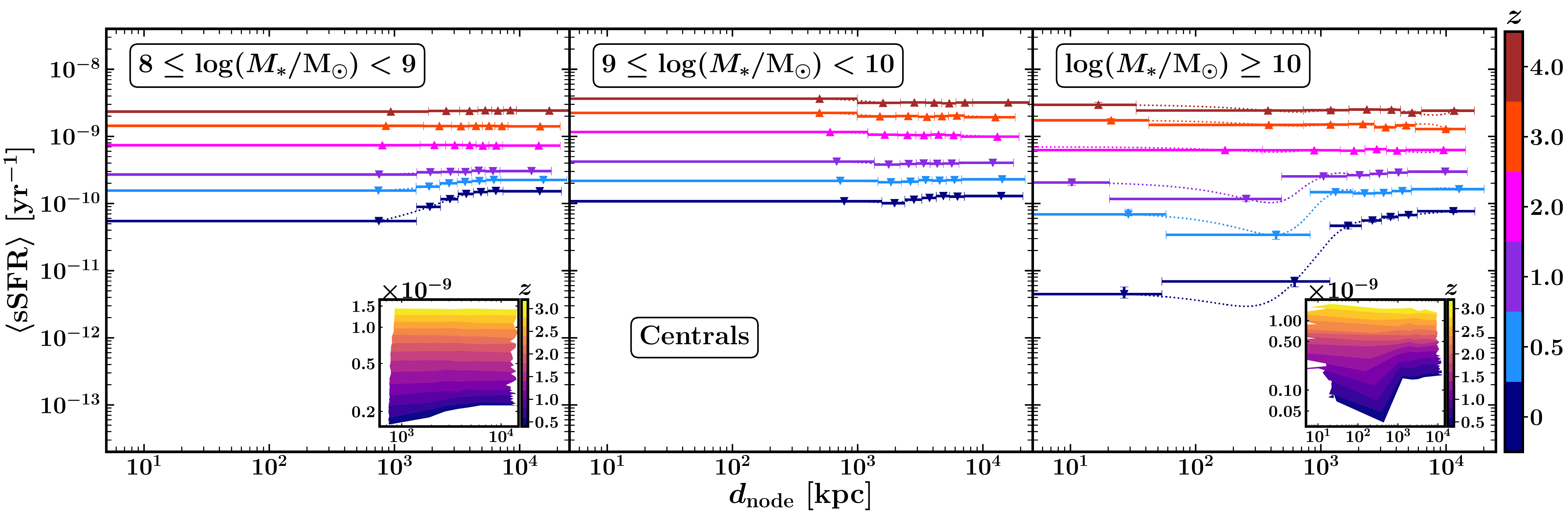}{0.925\textwidth}{}
} \vspace{-28pt}
\gridline{\fig{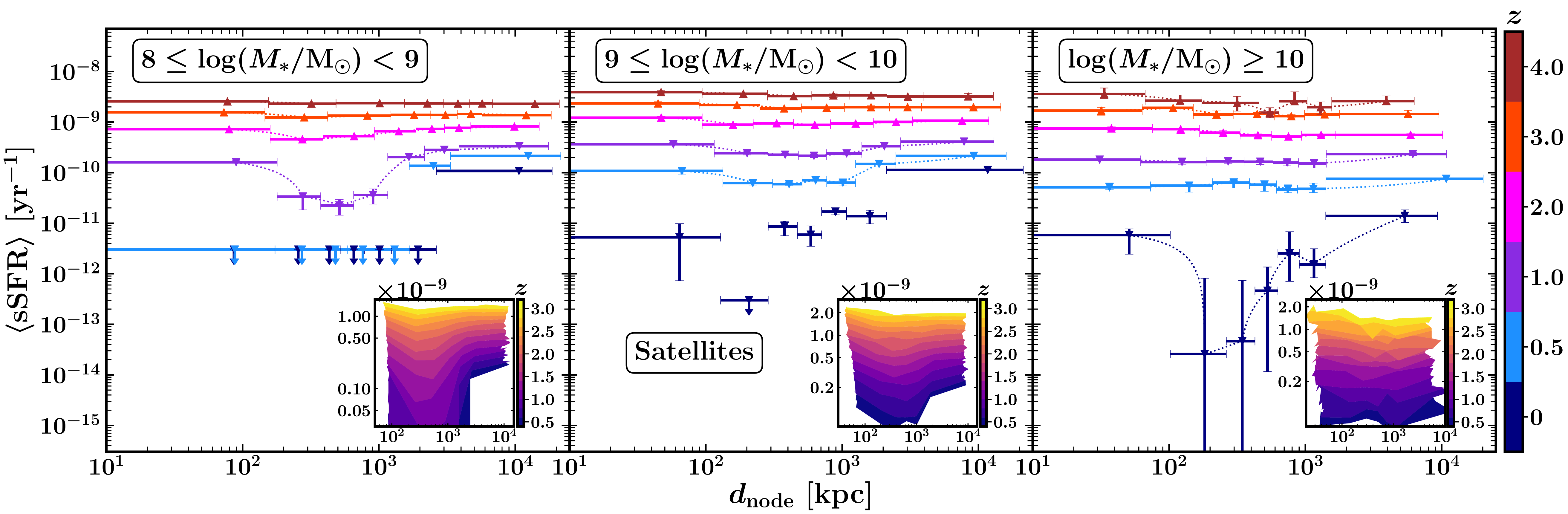}{0.925\textwidth}{}
} \vspace{-28pt}
\gridline{\fig{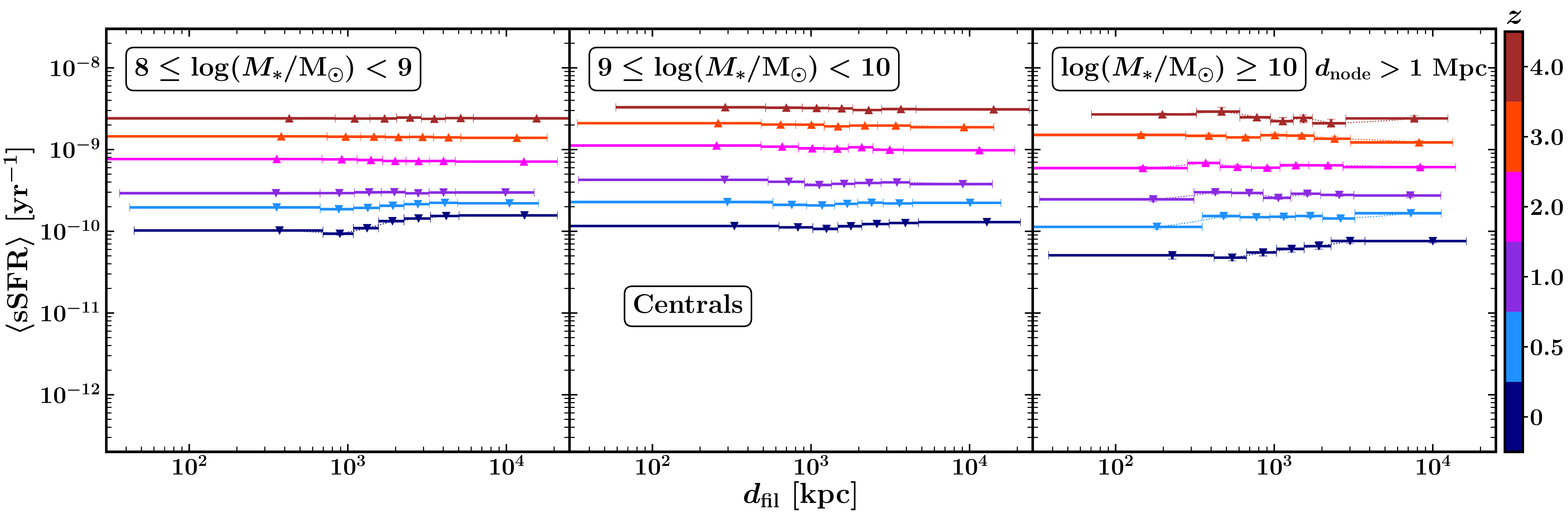}{0.925\textwidth}{}
} \vspace{-28pt}
\gridline{\fig{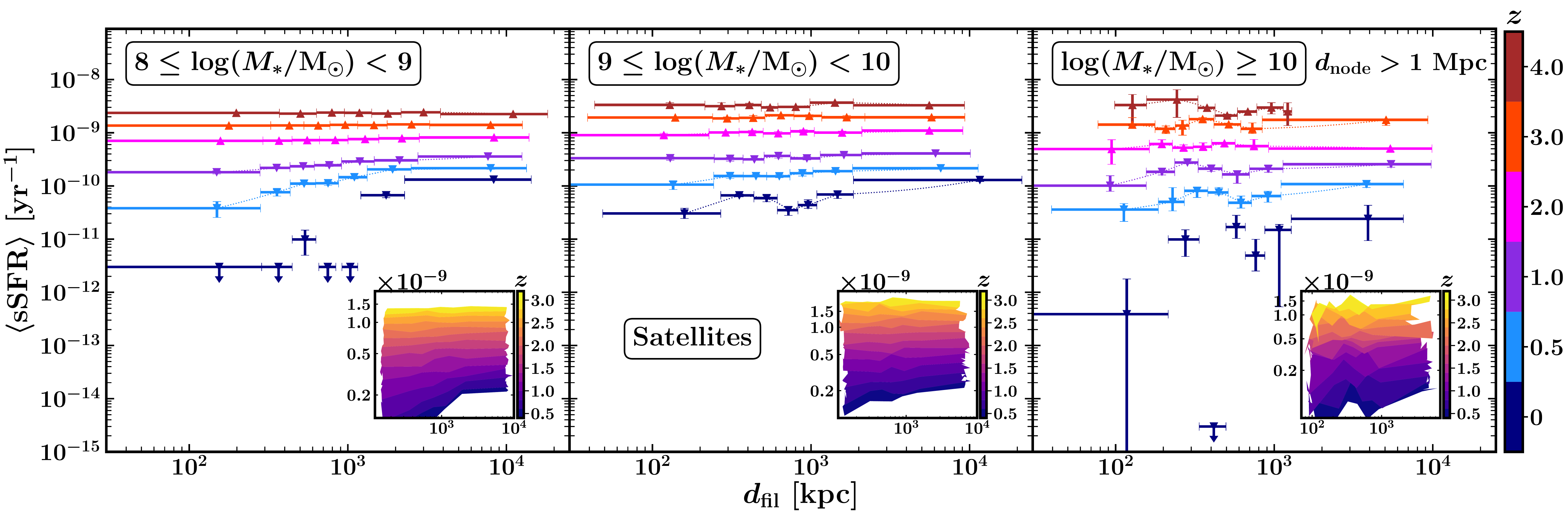}{0.925\textwidth}{}
} \vspace{-25pt}
\caption{
{\medssfr} in bins of {\dnode} (top two rows) and {\dfil} (bottom two rows) for central galaxies (\nth{1} and \nth{3} row) and satellite galaxies (\nth{2} and \nth{4} row) at different redshifts. Star formation in central galaxies is less dependent on cosmic web environment than that in satellite galaxies which are significantly quenched at low {\dnode} and {\dfil} at low redshifts. Neither centrals nor satellites exhibit a cosmic web dependence of star formation activity at $z\geq2$.  Insets are not included where no significant
relationships between cosmic web environment and sSFR are 
seen. 
}
\label{fig:centralvsat}
\end{figure*}

Considering the filament-centric relationships, we find both differences and similarities with the node-centric relationships. For low-mass galaxies at $z=0$, there is a sizeable $\sim5\times$ increase in {\medssfr} from $\langle {\dfil} \rangle \!\sim\! 0.3$~Mpc to $\langle {\dfil} \rangle \!\sim\! 15$~Mpc, however, the rise in {\medssfr} with {\dfil} is much smaller at $z=0$ ($\lesssim2\times$) and negligible at $z\geq1$. For the intermediate mass range, there is effectively no gradient of {\medssfr} with {\dfil} at any redshift. For high-mass galaxies at $z=0$, we do not see an upturn in {\medssfr} at the lowest {\dfil} (unlike at low {\dnode}) but rather a somewhat smooth rise by a factor of a few times in {\medssfr} with {\dfil} from low to high {\dfil}. At $z>0.5$, the relationship between {\medssfr} and {\dfil} is very weak. Thus, only low-mass and high-mass galaxies at low redshifts are preferentially quenched near filaments, whereas galaxies of all masses are impacted near nodes.

One of the most striking findings on the star formation-cosmic web connection, seen in both the filament- and node-centric analyses, is the {\it disappearance of a dependence of star formation on distance to cosmic web structures at higher redshifts}. 
The color contour insets included in Fig.~\ref{fig:medssfr} show the {\medssfr}-{\dnode} and {\medssfr}-{\dfil} relationships for many snapshots at $0.5 \!\leq\! z \!\leq\! 3$. 
These contours flatten out past a certain redshift for all three mass ranges, indicating that star formation activity is essentially independent of proximity to the cosmic web prior to this epoch.
The independence of star formation on cosmic web node-centric distance occurs at $z\sim1.3$ for {\midmsrange} galaxies and $z\sim2$ for {\lowmsrange} and {\highmsrange} galaxies. The {\dfil}-independence of star formation occurs at $z\sim1$ for low-mass and high-mass galaxies, while the star formation in moderate-mass galaxies does not show any significant {\dfil}-dependence at any redshift.
From our analysis, it can be deduced that the cosmic web environment began affecting star formation activity during the latter stages of, or immediately after, the so-called ``cosmic noon'' of star formation when the star formation rate density in the universe peaked \citep[ending around $z\sim1.5$; e.g.,][]{MD14}.


\subsection{Central And Satellite Galaxies}
\label{sec:satvcen}

We separate our galaxy samples into central and satellite galaxies to investigate how star formation depends on cosmic web environment for both galaxy types. At each redshift, we identify the most massive galaxy in a halo as the central galaxy and the rest as satellite galaxies and then repeat the analysis in Section~\ref{sec:sf}. 
The {\medssfr}-{\dnode} and {\medssfr}-{\dfil} relationships are shown for central and satellite galaxies in Fig.~\ref{fig:centralvsat}. 
We only include the color contour inset plots in the panels where any significant relationship between cosmic web environment and star formation is seen.

In general, we find that at low redshifts, the star formation in satellite galaxies is much more strongly connected to cosmic web environment than that of central galaxies. 
At $z\!<\!1$, there is a very modest rise in {\medssfr} of low-mass centrals with {\dnode} and in intermediate-mass centrals, there is no dependence of {\medssfr} on {\dnode}. 
In contrast, star formation is effectively quenched close to nodes in low-mass satellites at $z\!\leq\!0.5$, and in intermediate-mass satellites at $z\!=\!0$.

High-mass centrals show a strong correlation between {\medssfr} and {\dnode} at low redshifts. While the rise in {\medssfr} with {\dnode} is mostly monotonic at $z\!=\!0$, we see an upturn in {\medssfr} at low {\dnode} (${\dnode}\!\lesssim\!0.1$~Mpc) at $z\!\sim\!0.5-1$, similar to that found for the full galaxy population at the lowest redshifts (Fig.~\ref{fig:medssfr}). This upturn is also seen in high-mass satellites at $z\!=\!0$, suggesting that the elevation of star formation activity of high-mass galaxies very close to nodes is applicable for both centrals and satellites (with the caveat that the errors for the satellite relationships are larger). Beyond ${\dnode}\!\sim\!0.1$~Mpc, there is a smooth rise in {\medssfr} with {\dnode} for satellites at $z\!\leq\!0.5$ and for centrals at $z\!\leq\!1$.

With respect to filaments, there is negligible dependence of {\medssfr} on {\dfil} in centrals of any mass across cosmic time. Both low-mass and high-mass satellites are effectively quenched at low {\dfil} at $z\!=\!0$, while the rise in {\medssfr} with {\dfil} is more modest in intermediate-mass satellites. 
However the small-number statistics of the high-mass satellite population prevents us from drawing strong conclusions about their star formation dependence. 
While satellites appear to drive much of the dependence of star formation on proximity to cosmic web filaments and nodes at low redshifts, there is no statistically significant dependence of star formation on cosmic web environment at $z\!\geq\!2$ for either centrals or satellites.


\subsection{Gas Fraction And Cosmic Web Environment}
\label{sec:fgas}

\begin{figure*}[!hbt]
\vspace{-10pt}
\centering
\gridline{\fig{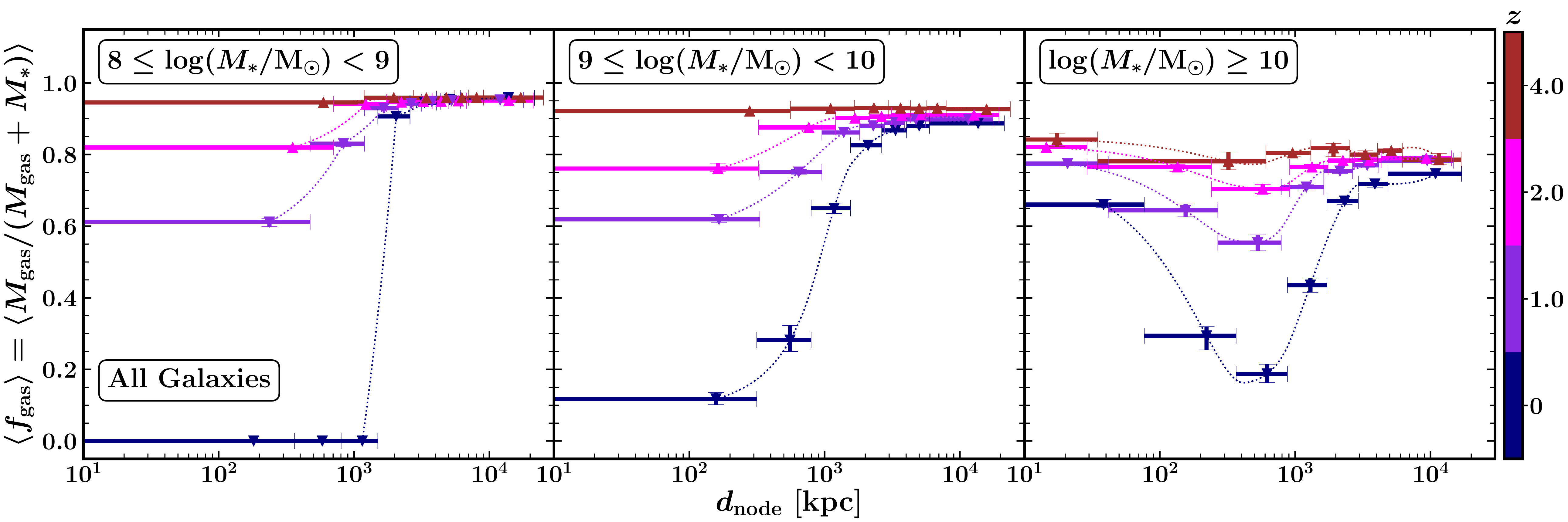}{0.85\textwidth}{}
} \vspace{-28pt}
\gridline{\fig{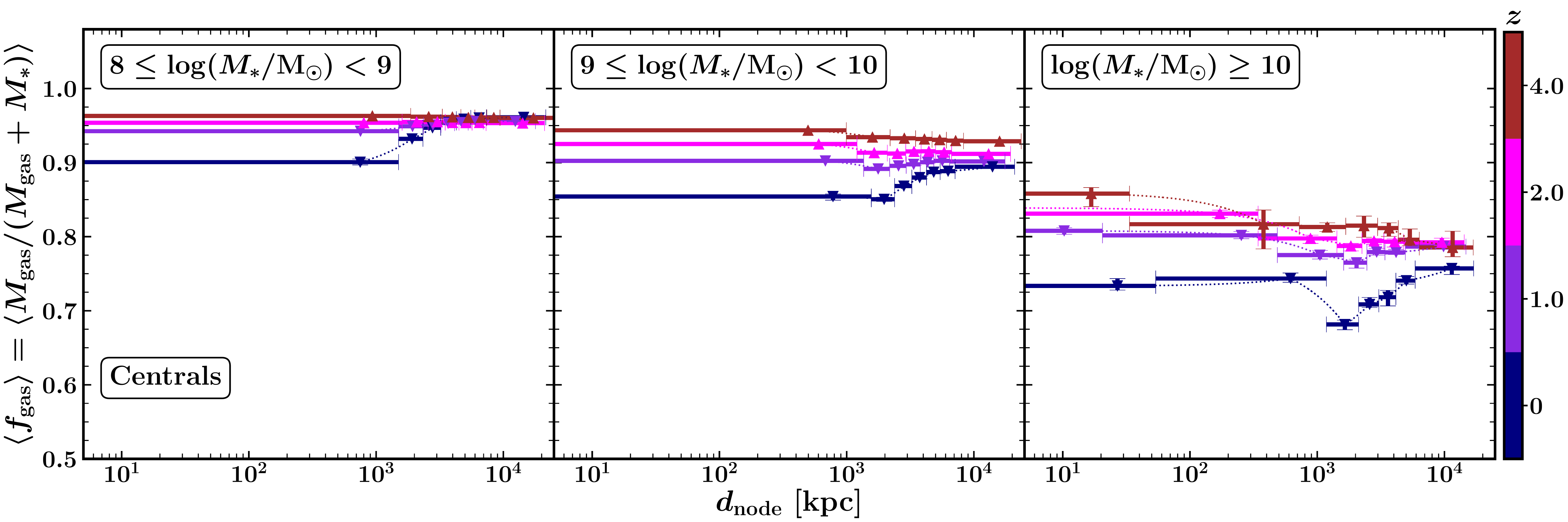}{0.85\textwidth}{}
} \vspace{-28pt}
\gridline{\fig{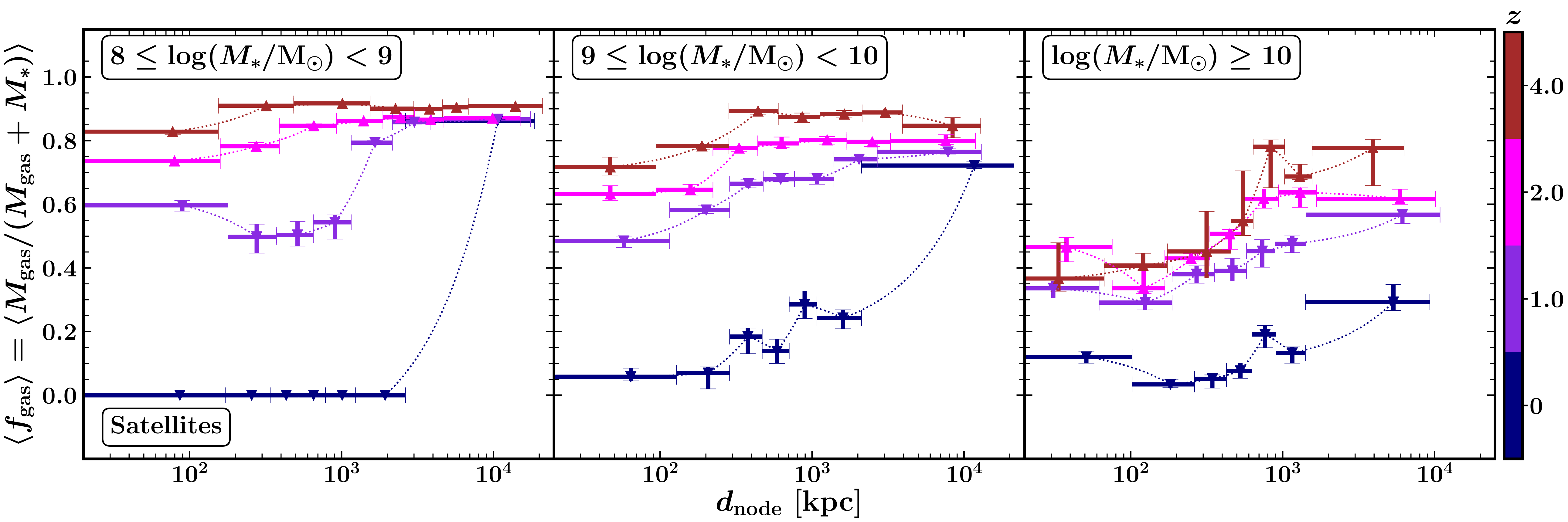}{0.85\textwidth}{}
} \vspace{-25pt}
\caption{
The median gas fraction, {\medfgas}, as a function of {\dnode}, for all galaxies (top row), centrals (middle row), and satellites (bottom row), at redshifts $z=0$, 1, 2, and 4. 
The general cosmic web dependence of star formation follows from the available gas supply. Satellites drive the gas fraction trends for lower mass galaxies while both centrals and satellites drive the high-mass trends.
}
\label{fig:fgas_dnode}
\end{figure*}

\begin{figure*}
\vspace{-10pt}
\centering
\gridline{\fig{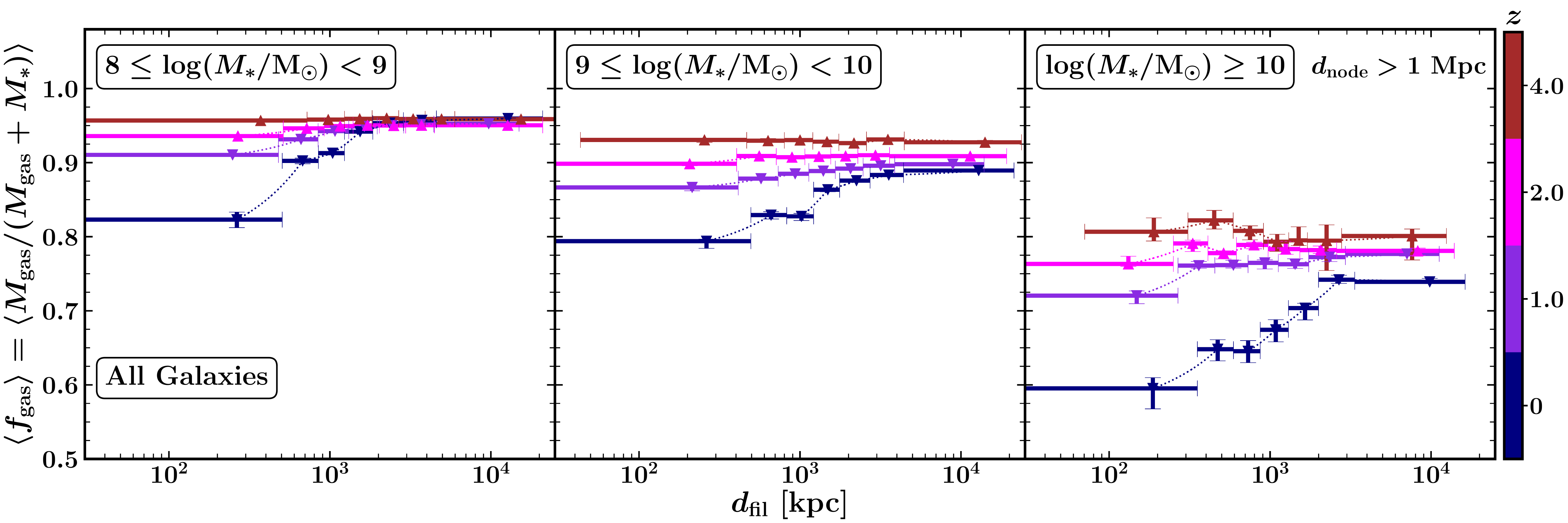}{0.85\textwidth}{}
} \vspace{-28pt}
\gridline{\fig{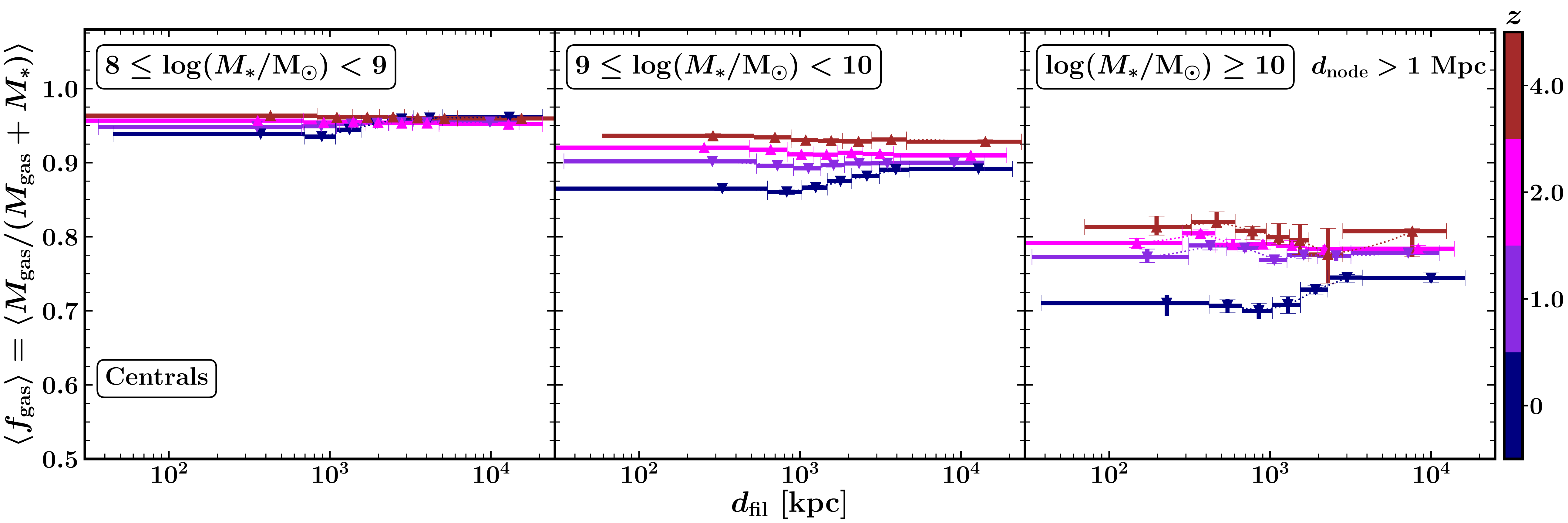}{0.85\textwidth}{}
} \vspace{-28pt}
\gridline{\fig{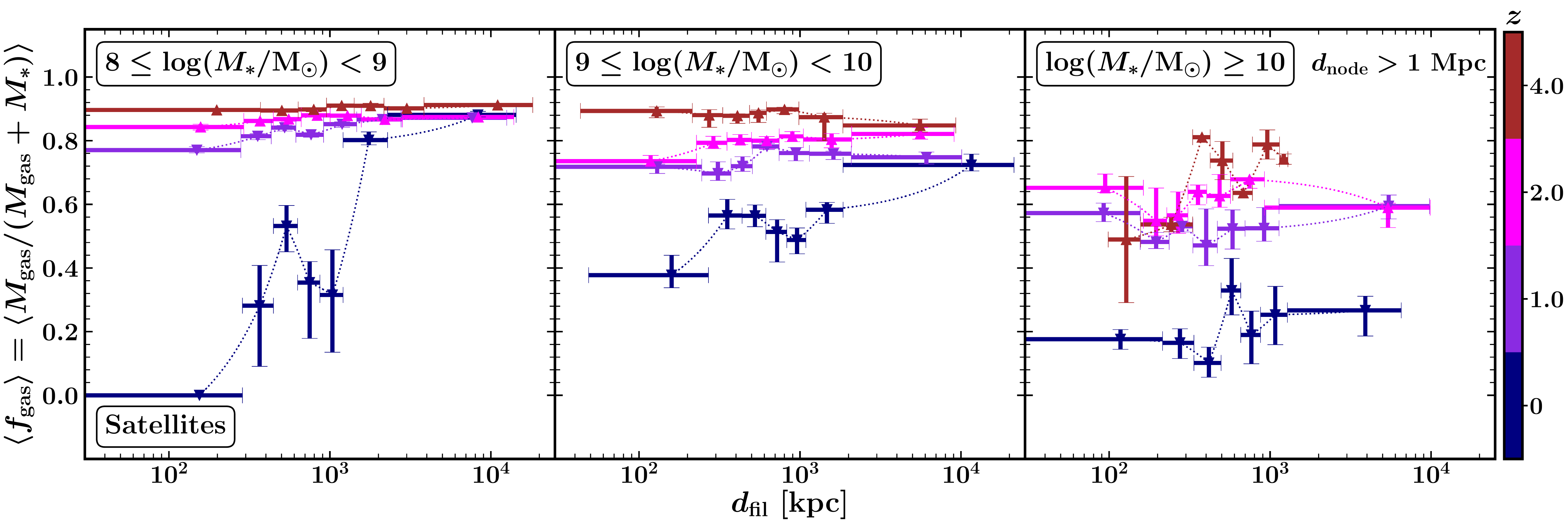}{0.85\textwidth}{}
} \vspace{-25pt}
\caption{
Same as Fig.~\ref{fig:fgas_dnode}, but in bins of {\dfil}. Satellites drive the dependence of {\medfgas} on {\dfil} more than centrals.}
\label{fig:fgas_dfil}
\end{figure*}

To further investigate the star formation-cosmic web connection, we examine the available gas content in galaxies relative to nodes and filaments. To this end, we measure the gas fraction,
\begin{equation} 
{\fgas} 
= \frac{ {\mgas} }{{\mgas} + {\ms}}\, ,
\label{eq:fgasgal}
\vspace{-5pt}
\end{equation}
which is the ratio of gas mass to the sum of gas and stellar mass bound to a galaxy. We calculate the median gas fraction, {\medfgas}, for all galaxies, centrals and satellites, in the same bins of {\dnode} and {\dfil} for the same mass ranges and redshifts as above. 
The {\medfgas}-{\dnode} and {\medfgas}-{\dfil} relationships are presented in Figures \ref{fig:fgas_dnode} and \ref{fig:fgas_dfil}, respectively. 
For clarity of presentation, we only include four redshift bins, $z=0,1,2,$ and $4$ in these plots. As in Fig.~\ref{fig:medssfr}, the $\pm1\sigma$ bootstrapped error-bars on the medians are shown.

Fig.~\ref{fig:fgas_dnode} shows that for the full galaxy population, there is a strong dependence of {\medfgas} on {\dnode} at lower redshifts. At $z\!=\!0$, low-mass galaxies at ${\dnode}\!\lesssim\!1$~Mpc are completely devoid of gas, which explains why they are quenched in these environments. This is mostly driven by satellite galaxies, as low-mass centrals only see a drop of a few percent in {\medfgas} from higher to lower {\dnode}. In intermediate-mass galaxies, {\medfgas} drops by an order of magnitude from the highest to the lowest {\dnode} bin at $z\!=\!0$, which is commensurate with the $\sim$1 dex drop in {\medssfr} in Fig.~\ref{fig:medssfr}. This is again primarily driven by a dramatic decline in gas fraction of satellites while centrals only exhibit a modest decline.

In high-mass galaxies, we find a minimum in {\medfgas} at ${\dnode}\!\approx\!0.7$~Mpc followed by a steep rise at lower {\dnode}. This dramatic upturn in gas fraction helps explain the upturn in star formation at low {\dnode} at low redshifts, but this turnover is a) much more pronounced in {\medfgas} than in {\medssfr} and b) persists out to $z=4$ for {\medfgas} while it only exists out to $z\sim1$ for {\medssfr}. The relationship between {\medfgas} and {\dnode} in high-mass galaxies is largely dictated by central galaxies, which have large gas fractions close to nodes across cosmic time. This, however, does not result in highly enhanced star formation in low-redshift centrals at low {\dnode}, possibly implying a lack of star formation efficiency in these environments  -- possibly due to central AGN heating the gas in these galaxies and suppressing star formation (see discussion below).
High-mass satellites also exhibit a small upturn in {\medfgas} at ${\dnode}\!\lesssim\!0.1$~Mpc at $z\!\leq\!3$. 
Unlike with star formation activity, the gas fraction in all {\highmsrange} galaxies depends on the proximity to nodes even at $z\!=\!4$.

Proximity to filaments is less strongly correlated with {\medfgas} than proximity to nodes, as shown in Fig.~\ref{fig:fgas_dfil}. At low redshifts, a monotonic rise in {\medfgas} with {\dfil} in low and intermediate-mass galaxies is caused mostly by the steep rise in the satellite population, but this dependence disappears with increasing redshift.
Centrals of all masses show virtually no {\dfil}-dependence on {\medfgas} with at any redshift, consistent with a lack of dependence of star formation activity on distance to filaments. 

We also note that the declining availability of gas in galaxies should not immediately result in reduced star formation activity; instead, there should be a time lag between the reduction in gas and the reduction in star formation. This is generally consistent with our results: {\medfgas} decreases at small {\dnode} and {\dfil} at higher redshift than does {\medssfr} for any given stellar mass range. In satellites and even high-mass centrals, a correlation between {\medfgas} and {\dnode} exists out to $z\!=\!4$ while star formation is independent of {\dnode} at $z\!\geq\!2$. 
Alternatively, these results can also be explained by star formation being less efficient further from nodes than closer to nodes at earlier times.

Our results in this section suggest the following. 1) Star formation quenching near cosmic web structures is typically preceded by a scarcity of gas, 2) the gas fraction in satellite galaxies is more significantly affected by cosmic web environment than that in central galaxies and drive the general cosmic web dependence of gas fraction, and 3) at later times, high-mass galaxies, including both satellites and centrals, are more gas-rich near centers of nodes than at the outskirts, leading to increased star formation closer to nodes.


\section{Discussion}
\label{sec:discuss}

\subsection{Physical Interpretations}
\label{sec:physical}

Here, we interpret our results in terms of physical mechanisms governing the evolution of galaxies and large-scale structure. 
Perhaps the most puzzling result is that the star formation activity of galaxies in TNG does not depend on their large-scale cosmic web environment at $z\!\geq\!2$, in contrast with later times when significant dependence occurs.
At face value, this seems to suggest a rising importance of quenching driven by environment with cosmic time and, indeed, several recent works have found varying degrees of evidence for a lack of small or large-scale environmental dependence of star formation at higher redshifts -- in both observations \citep[e.g.,][]{Moutard18,Chang22,Momose22} and simulations \citep[e.g.,][]{Xu20}.


\subsubsection{The Cosmic web dependence After Cosmic Noon}

The low-redshift dependence of star formation activity on {\dnode} and {\dfil} can generally be explained by the variation of gas fraction with these distances. In lower mass galaxies at low redshifts, the monotonic descent in {\medssfr} towards nodes and filaments is consistent with the corresponding descent in {\medfgas}. On average, low-mass galaxies that are within several hundred kpc of a node or filament effectively stop forming new stars at $z\!=\!0$, most likely because they have very little to no gas available to do so, 
a behavior largely driven by satellites.

This points to a picture where dwarf galaxies that accrete onto the halos of more massive centrals are quenched in overdense environments where their gas supplies are depleted. 
Dwarf satellites located in galaxy clusters and groups are subjected to harsh gaseous environments dominated by warm and hot gas with high densities and long cooling times.
In these environments, a combination of physical processes can act together on the gas reservoirs of dwarf satellites.

Gas may be removed by ram pressure stripping when these galaxies move through the group/cluster medium or by gravitational (tidal) interactions between the satellites and the central/other satellites or the halo itself. These processes, while often identified as the likely culprits for gas stripping and quenching of low-mass satellites in clusters/groups, typically act on relatively short timescales of $\lesssim$500~Myr \citep[e.g.,][]{BM15,Marasco16}. Here, we find that low-mass satellites close to nodes are already quenched at $z\!=\!0.5$ and have greatly reduced gas fractions at $z\!=\!1$, meaning that the quenched satellites in the local universe had been quenched much earlier. In fact, \citet{Donnari21a} found that in TNG, a large fraction of $z\!=\!0$ $\log(\ms/{\Msun})\!\lesssim\!10$  satellites in groups and clusters were members of other halos whence they experienced environmental quenching before falling into their final host -- a phenomenon dubbed ``pre-processing'' \citep[see also, e.g.,][]{Fujita04,Hou14}.

AGN feedback from massive central galaxies in groups and clusters may also play an important role in quenching star formation in satellites. The TNG model allows for both ``ejective'' feedback whereby BHs expel star-forming gas from a galaxy and ``preventative'' feedback whereby BHs heat up the gas and prevent star formation on longer timescales \citep{Zinger20}. 
Both these modes of AGN feedback have been observed in galaxies near and far \citep[e.g.,][]{Fabian12,KP15}. In particular, \citet{MN19} showed that stronger BH feedback produces hotter group/cluster media that makes quenching more efficient in satellites. While beyond the scope of this work, it would be valuable to understand how central BH properties such as mass and accretion rate might relate to the cosmic web dependence of star formation.

Satellites at close filament-centric distances and ${\dnode}\!>\!1$ Mpc would reside in more intermediate density environments than those at close node-centric distances. In either of these types of environments, fresh gas accretion onto the ISM can be stopped by strangulation/starvation such that star formation quenches over longer timescales of $\sim$few Gyr \citep[e.g.,][]{Peng15,Zinger18}. In a comprehensive analysis of nearby galaxies, \citet{Trussler20} found that starvation is likely to be the initial prerequisite for quenching across virtually all masses but the remaining cold ISM gas needs to be heated or ejected to complete the quenching process. 
Low-mass centrals, on the other hand, exhibit a modest cosmic web dependence on gas fraction and consequently, star formation. For these galaxies, so-called ``mass quenching'' via internal processes (such as feedback) may dominate over environmental effects \citep[e.g.,][]{Peng10}.

In our investigation, we considered the total content of all gas gravitationally bound to a galaxy, without regards to the physical conditions or location of the gas.
For instance, we did not measure the fraction of {\it cool} gas which would in principle be a more direct measure of gas supply available for star formation. 
Regardless, we find that the gas fraction of low-mass satellites declines dramatically from $z\!=\!1$ to $z\!=\!0$ at small {\dnode} and to a lesser, but still significant, extent at small {\dfil}, indicating a lack of accretion from the IGM to the CGM. Many hydrodynamical simulations indeed show that accretion of cold gas from the IGM becomes increasingly inefficient over time \citep[e.g.,][]{AA17,Hafen20}. Furthermore, gas in the CGM may be heated substantially or even ejected by SNe \citep[e.g.,][]{Pandya22} or AGN (as discussed above) to prevent accretion onto the ISM.


In low-redshift {\highmsrange} galaxies, a minimum in gas fraction and star formation activity occurs at ${{\dnode}\!\sim\!0.2}$~Mpc, following an unexpected rise at smaller {\dnode}. 
This upturn in {\fgas} and sSFR close to nodes persists out to $z\!=\!2$, but manifests in an analogous relationship between star formation and proximity to nodes at $z\!\leq\!1$. 
When we limit our sample to even higher mass galaxies, this effect is further accentuated, implying that the highest mass galaxies are primarily responsible.
The effect of enhanced star formation very close to nodes is stronger in satellites than in centrals at $z\!=\!0$ while the converse is true at $z\!=\!0.5-1$. The fact that both {\medssfr} and {\medfgas} are lowest at ${\dnode}\!\sim\!0.2$~Mpc implies that massive galaxies falling into groups/clusters from the outskirts are more gas-poor and passive relative to galaxies at the center.

It is conceivable that some/much of the gas removed from low-mass satellites in rich groups and clusters ends up in higher mass galaxies, enabling the massive galaxies to form stars at higher rates near the centers of these halos. The cores of many groups and clusters have been observed to be abundant in cold gas, which may temporarily trigger star formation near the center \citep[e.g.,][]{McDonald12,Olivares19}. However, this gas is also hypothesized to feed central AGN activity and eventually curtail star formation \citep[see][and references therein]{DV22}. Heating from the central AGN could explain why the rise in star formation at small {\dnode} is not as dramatic as the rise in gas fraction in high-mass galaxies.

The enhancement of star formation in dense environments is not typically observed in statistical studies of the cosmic web-galaxy connection \citep[e.g.,][]{Kraljic18, Winkel21}. But there is evidence -- both in observations \citep[e.g.,][]{Roediger14} and in simulations \citep[e.g.,][]{Nelson18} -- of galaxies in groups/clusters enjoying brief episodes of star formation via compression of gas from ram pressure, mergers or other processes, a phenomenon sometimes called ``rejuvenation.'' However, these events are rare in TNG, with only $10\%$ of $\log({\ms}/{\Msun})\!>\!11$ galaxies and $6\%$ of all galaxies at $z\!=\!0$ ever having experienced them \citep{Nelson18}. 
An analysis of satellite galaxies by \citet{MN21} showed that AGN outflows can clear out the CGM of massive halos, which reduces ram pressure and preserves star formation in satellites along the direction of the outflows (the minor axis of the central). These phenomena of {\it positive} AGN feedback can potentially boost star formation in dense environments such as those close to nodes.

It is also possible that the high-density upturn in star formation is a result of some additional mechanism in the simulations funnelling too much gas into galaxies, over-cooling the gas, or otherwise reducing the efficiency of quenching at the highest density environments. 
\citet{Donnari21a} 
report that TNG galaxies in dense environments have diverse histories and quenching pathways that may complicate the interpretation of how and when they quench. Moreover, the AGN-driven gas expulsion in TNG is known to be so efficient that there are very few galaxies with intermediate sSFR \citep[i.e., Green Valley galaxies; e.g.,][]{Sch14}, creating tension with observations 
\citep[][]{Terrazas20}.



\subsubsection{No Cosmic web dependence Before Cosmic Noon?}

There is now a growing body of evidence suggesting that cosmological accretion shocks from the formation of cosmic web structures, similar to those around massive halos, can affect galaxy formation.
\citet{Birnboim16} showed that in filaments with a specific linear mass density, the accretion shocks are unstable. These structures can efficiently siphon cool gas into $10\!\lesssim\!\log({\mh}/{\Msun}) \!\lesssim\!13$ halos at $z=3$ and 
$12 \!\lesssim\!\log({\mh}/{\Msun})\!\lesssim\!15$ halos at $z\!=\!0$ (see their Fig. 5). According to the stellar-to-halo-mass relations at these redshifts \citep[e.g.,][]
{Behroozi19}, this means that unstable filaments can potentially enhance star formation in galaxies of virtually all masses we studied at higher redshifts, while at lower redshifts only the most massive galaxies would see an increase in star formation via this channel. This phenomenon is a possible pathway for early galaxies close to filaments and nodes to have their sSFR elevated to levels comparable to those far from filaments and nodes.

This explanation necessitates an environmental dependence of {\it overall star formation activity} at high $z$ instead of specifically {\it quenching}. The net trend of constant sSFR with distance from the cosmic web could be naively interpreted as quenching processes being environment-independent at high $z$. 
As noted in Section~\ref{sec:fgas}, we find evidence of star formation in satellites close to nodes being more efficient than that in galaxies further away at early times ($z\!\geq\!2$). A plausible scenario for this is that gas is more efficiently channelled into the centers of nodes, and eventually galaxies, via cold streams at high redshift \citep[e.g.,][]{Dekel09}.

Cosmological accretion shocks can also suppress star formation in galaxies. 
\citet{Zinger18} showed that accretion shocks at the outskirts of galaxy clusters can quench satellites, which likely impacts galaxies near nodes in our analysis.
In TNG, \citet{Li23} found that shock-induced stripping of the ISM and CGM can quench low-mass satellites inside clusters at $z\!<\!0.11$.
Recently, \citet{Pasha22} found that $5.5 \!<\! \log({\ms}/{\Msun}) \!<\! 8.5$ central galaxies at $z\!=\!2-5$ can be quenched by shock-heated cosmic sheets (which eventually collapse into filaments and nodes; e.g., \cite{Bond96}). These shocks
directly raise the ambient gas temperature in the vicinity of the sheets and suppress gas accretion and star formation in surrounding galaxies.

The impact of accretion shocks in filaments and nodes on galaxy quenching, as a function of both stellar mass and redshift, may be central to interpreting the results of this paper and therefore deserves detailed investigation. In a follow-up study, we will address this problem by analyzing the gaseous conditions of filaments and nodes -- with particular emphasis on accretion shock signatures -- in tandem with properties of the galaxies residing within them across cosmic time. 
This analysis will also allow us to characterize filaments and nodes in more detail and account for the fact that not all filaments or nodes will have the same effect on galaxy formation (e.g., \citealt{GE20a} found short and long filaments in TNG to be statistically different populations).


\subsubsection{Other Important Physical Considerations}

Angular momentum is another important aspect of galaxy formation which may shed additional light on how quenching is affected by the cosmic web. 
From their analysis of quenching timescales in TNG, \citet{Walters22} suggested that low angular momentum gas accretion leads to galaxies quenching faster than high angular momentum accretion.
In the cosmic web framework, galaxies form in the vorticity-rich regions of filaments, acquire angular momentum, and drift to the nodes \citep[e.g.,][]{Dubois14,Codis15}. 
Simulations have predicted for many years that at $z\!\gtrsim\!1.5$, gas and angular momentum are funnelled through cold filamentary streams into the centers of galaxies 
\citep[e.g.,][]{Dekel09,Pichon11}. Over time, as these streams disappear due to heating or other processes, the efficiency of galaxy formation at the centers of filaments and nodes may also decline, potentially explaining the difference in star formation activity in these regions between low and high redshift.
Additionally, galactic properties such as mass and sSFR have been found to be correlated with the acquisition of angular momentum from the cosmic web \citep[e.g.,][]{Kraljic19,Welker20}. 
Thus, a complete understanding of how the cosmic web affects quenching needs to account for angular momentum acquisition in galaxies in tandem with proximity to cosmic web structures.

Many of the relationships between star formation and cosmic web environment may result from the assembly of DM halos. 
Subhalo abundance matching predictions from $\Lambda$CDM cosmology are found to agree with observed SDSS galaxy distributions, implying that the local density dependence of galaxy properties stems from the corresponding density dependence of halo properties \citep{Dragomir18}. 
In the Bolshoi-Planck simulations, \citet{Lee17} found that for $\log({\mh}/{\Msun})\lesssim12$ halos, halo accretion is higher in low-density environments at $z\lesssim1$ and in high-density environments at $z\!\gtrsim\!1$. 
However, some $N$-body simulations have shown that DM halo properties are independent of cosmic web location at fixed overdensities \citep[e.g.,][]{Goh19}. 
A detailed analysis of the dependence of halo mass accretion with cosmic web environment in TNG is necessary to disentangle the effect of halo mass growth from baryonic effects in determining the galaxy quenching-cosmic web connection.


\subsection{Other Caveats}

We consider certain other aspects of our methodology that may affect the robustness of our results as well as the conclusions we draw. The first is how numerical resolution in the simulation may affect our results. \citet{GE21} showed that despite the $\sim$8 times difference in resolution between TNG300-1 and TNG300-2, there are only minor differences in the distribution (including the {\disperse} reconstruction) and properties of filaments. The large scales of the cosmic web are likely to be well-resolved with any of the TNG runs, but the smaller scales of galaxy formation are more sensitive to resolution. Thus, it would be interesting to compare our results for TNG100-1 with those of TNG50-1, which has $\sim$16 times the particle mass resolution of TNG100-1 \citep[e.g.,][]{Nelson19,Nelson20}.

The input physics model is another potential source of uncertainty for theoretical galaxy evolution studies. \citet{GE20a} found that different baryonic physics implemented in different simulations result in somewhat different matter distribution around filaments but that gravity is still the dominant driver. \citet{Xu20} investigated the sSFR of galaxies in filaments, nodes, sheets, and voids in the EAGLE simulation which uses somewhat different hydrodynamics and feedback prescriptions from the TNG model \citep[see][]{Schaye15}. They found that at $z\!<\!1$, galaxies with $\log({\ms}/{\Msun})\!\lesssim\!10.5$ are less star-forming in nodes than other cosmic web environments, while for more massive galaxies there is virtually no cosmic web dependence, the latter finding being at odds with our results. At $z\!>\!1$, they found no statistical dependence of sSFR on the cosmic web environment, consistent with our findings. In a separate study, \citet{RG22} found that the star-forming fraction of $\log({\ms}/{\Msun})\!>\!9$ galaxies in EAGLE decreases with distance to the nearest void at $z\!=\!0$. It would be interesting to apply our methodology to investigate the evolving cosmic web dependence of star formation in other hydrodynamical cosmological simulations such as SIMBA \citep{Dave19} and Horizon-AGN \citep{Dubois14}. 
Such comparisons might help illuminate the effect of uncertain baryonic processes such as AGN feedback on the relationship between galaxy formation and the cosmic web.

Another crucial check on our results is the cosmic web reconstruction itself. As mentioned in Section~\ref{sec:disperse}, we experimented with {\disperse} parameter choices such as persistence and smoothing of the filamentary skeleton. The latter did not have any significant effect on our results and varying the former affected the frequency of identified structures but did not affect any qualitative conclusions. Overall, we consider our results to be robust to parameter choices. There are several other cosmic web reconstruction techniques that have been employed for cosmic web studies, which have advantages and disadvantages over the {\disperse} framework \citep[see][for a detailed comparison of many of these methods]{Libeskind18}. 
We are currently applying a new state-of-the-art cosmic web reconstruction algorithm called the Monte Carlo Physarum Machine (MCPM), inspired by the {\it Physarum Polycephalum} (slime mold) organism \citep{Elek21,Elek22}, to compare to the local density estimation and global cosmic web characterization from {\disperse}. This method produces continuous cosmic matter densities (as opposed to discrete DTFE densities at the locations of galaxies) and has been applied successfully to both theoretical and observational datasets \citep[e.g.,][]{Burchett20,Simha20,Wilde23}.

We also assess the importance of local galaxy overdensity in shaping the star formation-cosmic web connection.
We repeat our analyses in Section~\ref{sec:results} by only considering galaxies with local DTFE galaxy overdensity (as computed by {\disperse}) within $\pm1\sigma$ of the mean. We find that the resulting relationships with respect to {\dnode} and {\dfil} look strikingly similar to those we report without filtering out galaxies by overdensity, implying that the effect of cosmic web environment on star formation persists beyond just the highest density regions of the universe.
However, we stress that local overdensity and global cosmic web environment are necessarily related to each other and it is therefore not trivial to disentangle the effects of one from the other.
We defer a detailed characterization of the dependence of {\dfil} and {\dnode} on overdensity across different redshifts in TNG100 to a future work \citep[see][for an in-depth mapping of overdensity to cosmic web proximity in TNG300]{Malavasi22}.

Finally, in our interpretations, we neglected the effect of pseudo-evolution of filaments and nodes, i.e., the evolution of the reference density (in our case, the DTFE mean density) instead of a true {\it physical} density. Such pseudo-evolution is known to strongly drive the mass-evolution in DM halos, especially at lower redshifts \citep{Diemer13}, and we defer a detailed investigation of this effect to future work.


\subsection{Testing Predictions with Observations}
\label{sec:obs}

The predictions presented herein from the TNG100 simulation establish clear objectives for observational studies.  First, we identified a point in cosmic time where galaxies' star formation activity begins to depend on their location relative to the large-scale cosmic web environment.  Confronting this prediction with observations will necessitate wide-field galaxy surveys capable of characterizing the large-scale structure over a large range of redshifts, out to at least $z\!=\!2$.  Second, a common theme we observed in both sSFR and gas fraction was an increase at small node-centric distances for high-mass galaxies. This will require both the extensive survey data necessary for finding, and perhaps even more challenging, measuring the gas contents of, these galaxies.  Spectroscopic galaxy surveys both underway and planned for the next several years should make serious headway towards at least the first element of this challenging observational experiment.

The current gold standard for wide-field spectroscopic surveys is SDSS, which can provide the lowest redshift anchor point for such a comparison.  The quoted SDSS spectroscopic completeness limit of $m_r\!=\!17.7$ would correspond to a redshift limit of $z\!\sim\!0.01$ for the lowest mass galaxies studied here ($10^8~ \Msun$). Even at $z\!\sim\!0.1$, SDSS is only complete to $\sim10^{10}~ \Msun$, covering the most massive bin we study.  Thus, SDSS, in principle, is capable of yielding measurements comparable with the dark blue data points in Figures \ref{fig:medssfr} and \ref{fig:centralvsat}.  Although an independent analysis of observational datasets with our methodology is beyond the scope of this paper, we refer the reader to the work of  \citet{Kuutma:2017aa}, \citet{CroneOdekon:2018aa}, and \citet{Winkel21}, who explore similar relationships with SDSS.  Also of note is \citet{Kraljic18} who employ the Galaxy and Mass Assembly (GAMA) survey, which goes two magnitudes deeper than SDSS, albeit over a much smaller volume.  Still, this does not push completeness to the $z\!>\!1$ transition point in cosmic web dependence we report here.

Constraining the higher redshifts will be more difficult, although the various Dark Energy Spectroscopic Instrument (DESI) surveys should enable cosmic web reconstructions at intermediate redshifts.  Initial data and results from the Survey Validation phase are beginning to be released now, showing promising prospects for mapping the large-scale structure to $z\!\sim\!0.5$ for the Bright Galaxy Survey, $z\!\sim\!1.1$ for the Luminous Red Galaxies (LRGs), $z\!\sim\!1.6$ for Emission Line Galaxies (ELGs), and possibly beyond \citep[][and references therein]{Lan:2023_desiVisual}.  However, each of these samples is likely to contain highly biased tracers of the underlying structure; e.g., the LRGs are by definition passive galaxies and will preferentially reside in the most massive halos, likely tracing nodes.  Conversely, the ELGs, being vigorously star-forming, might bias against these very environments.  Nevertheless, neither of these samples will suitably represent the full diversity in star formation exhibited by the general population.  Deep follow-up surveys with more agnostic selection criteria will be necessary for a fair comparison with our results.

The Subaru Prime Focus Spectrograph (PFS) also offers great promise for mapping out the galaxy-cosmic web connection at higher redshifts \citep{Takada14}. In particular, the PFS Galaxy Evolution program will observe up to half a million galaxies at redshifts $0.7 \!\lesssim\! z \!\lesssim\! 7$ \citep{Greene22}. The largest survey of this program is expected to yield $>\!10^{5}$ continuum-selected galaxies down to a stellar mass limit of $\log(\ms/\Msun)\!\approx\!10.5$ over a comoving survey volume of $\sim\!0.1~\mathrm{Gpc}^3$ ($\sim$100 times the TNG100 volume) at $0.7 \!\lesssim\! z \!\lesssim\! 2$. Stellar masses, SFRs, and gas properties will be measured for the vast majority of these galaxies. This sample will be complemented by a smaller number of LBGs and Lyman Alpha Emitters (LAE) out to $z\!\sim\!7$ which would be more biased tracers of the cosmic web as discussed above.

Spectroscopic surveys with \textit{JWST} and, eventually, the \textit{Nancy Grace Roman Space Telescope} will yield galaxy datasets ripe for placing in context with the cosmic web mapped by DESI and Subaru PFS. \textit{Roman}, which will map 1700 deg$^2$ of the sky at infrared wavelengths via the High Latitude Spectroscopic Survey \citep{Wang:2022_romanHLSS}, should reveal the cosmic web over scales of $\sim$1 Gpc as well as the galaxies within to $z\!\sim\!2$.  In the shorter term, \textit{JWST}, through programs such as JADES \citep{Cameron:2023_jadesISM}, will yield galaxy spectra to $z>5$ albeit over much smaller fields of view (the \textit{JWST} Micro-shutter Assembly will map scales $\sim1$ Mpc across in a single pointing).  An amalgamation of several such deep fields will be necessary to mitigate cosmic variance.



\section{Conclusion}
\label{sec:conclusion}

In this study, we investigated the IllustrisTNG simulations to understand how the star formation activity of galaxies depends on their cosmic web environment. We used all {\minms} galaxies to reconstruct the cosmic web in the TNG100 snapshots using the {\disperse} framework. We measured the median sSFR and median {\fgas} of galaxies as functions of distance to the nearest cosmic web node ({\dnode}) and filament spine ({\dfil}).
Our main results are as follows:
\begin{enumerate}
\itemsep0em
\item The {\medssfr} of galaxies at any mass only depends on {\dnode} or {\dfil} at redshifts $z\!\lesssim\!2$; {\it the median star formation is independent of cosmic web environment at $z\!\geq\!2$}. This holds true also for central and satellite galaxies separately. 
\item In {\mslow} galaxies, {\medssfr} increases monotonically with {\dnode} at $z\!\leq\!1$, with {\lowmsrange} galaxies being completely quenched at ${\dnode}\!<\!1$~Mpc at $z\!\leq\!0.5$. {\medssfr} has a shallower increase with {\dfil} at these redshifts. These trends are almost entirely driven by satellites.
\item In {\highmsrange} galaxies, the {\medssfr}-{\dnode} relationship inverts at ${\dnode}\!\lesssim\!0.2$~Mpc up to $z\!=\!1$, while the {\medssfr}-{\dfil} relation does not. The {\medssfr}-{\dnode} inversion is driven by both satellites and centrals, but the {\medssfr}-{\dfil} relationship is due to satellites. 
\item Most of these star formation-cosmic web relationships can be explained by the cosmic web dependence of gas fraction in galaxies, although there is evidence of {\medfgas} depending more strongly on cosmic web environment than {\medssfr} in some cases.
\end{enumerate}

Our results point to a picture where the influence of the cosmic web environment on quenching galaxies is first established at $z\!\sim\!2$.
In the last $\sim$10~Gyr, low-mass dwarf satellites are quenched by their star-forming gas supplies being depleted either on short timescales (e.g., via ram pressure stripping or outflows) or on longer timescales (e.g., via starvation), while star formation in low-mass centrals is far less affected by cosmic web environment.
At this epoch, high-mass galaxies at the centers of nodes are more gas-rich and star-forming than their counterparts at the outskirts, which could be due to temporary rejuvenation events, positive AGN feedback, and/or a consequence of the TNG model itself. 
In the earlier universe ($>$10~Gyr ago), cosmic web structures likely aided star formation more than they suppressed it, possibly via unstable filaments feeding cold gas to galaxies or cold streams efficiently funnelling initially high angular momentum gas to the central regions of filaments and nodes.

In a follow-up study, we will investigate how the gaseous physical conditions of filaments and nodes affect galaxy formation in TNG100, in particular how accretion shocks around filaments and nodes affect star formation.
Furthermore, we will compare the cosmic web reconstruction from {\disperse} with that from the novel 
MCPM algorithm \citep{Elek21,Elek22} to obtain more fine-grained insights into the global and local environmental dependence of star formation across cosmic time. 
The results of this work provide important predictions to test against ongoing large spectroscopic surveys such as SDSS and DESI, as well as those ongoing and planned with Subaru PFS, {\it JWST} and {\it Roman}.


\begin{acknowledgements}

We are very grateful to N. Luber and Z. Edwards for help with setting up {\disperse}.
We thank the anonymous referee for helpful comments that improved the quality of this manuscript.
We thank attendees of the 2022 Santa Cruz Galaxy Workshop and the 2023 KITP Cosmic Web Conference, including  F. van den Bosch, J. Woo, H. Aung, J. Powell, C. Pichon, U. Kuchner, C. Welker, S. Simha, K-G. Lee, and R. Momose, for stimulating and interesting conversations on this work. 
FH, JNB, and AA are supported by the National Science Foundation LEAPS-MPS award $\#2137452$. 
OE is supported by an incubator fellowship of the Open Source Program Office at UC Santa Cruz funded by the Alfred P. Sloan Foundation (G-2021-16957).
DN is supported by NSF (AST-2206055) and NASA (80NSSC22K0821 \& TM3-24007X) grants.

\end{acknowledgements}


\bibliographystyle{aasjournal_nikki}
{\tiny \bibliography{Refs}}

\end{document}